\documentclass[debug]{rmaa}

\title{Kinematic ages of central stars of planetary nebulae} 

\author{
  W. J. Maciel,\altaffilmark{1} 
  T. S. Rodrigues\altaffilmark{1}
  and R. D. D. Costa,\altaffilmark{1}}

\altaffiltext{1}{Instituto de Astronomia, Geof\'\i sica e Ci\^encias
                 Atmosf\'ericas, Universidade de S\~ao Paulo, Brazil.}

\shortauthor{Maciel, Rodrigues, \& Costa}
\shorttitle{Kinematic ages of CSPN}

\fulladdresses{
\item W. J. Maciel, T. S. Rodrigues and R. D. D. Costa: 
    Instituto de Astronomia, Geof\'\i sica e Ci\^encias
    Atmosf\'ericas, Universidade de S\~ao Paulo - 
    Rua do Mat\~ao 1226, CEP 05508-090, S\~ao Paulo SP, Brazil
    (maciel@astro.iag.usp.br, tsrodrigues@usp.br, 
    roberto@astro.iag.usp.br.)}

\listofauthors{W. J. Maciel, T. S. Rodrigues, \& R. D. D. Costa}
\indexauthor{Maciel, W. J.}
\indexauthor{Rodrigues, T. S.}
\indexauthor{Costa, R. D. D.}

\abstract{The age distribution of the central stars of planetary nebulae 
(CSPN) is estimated using two methods based on their kinematic properties.  
First, the expected rotation velocities of the nebulae at their 
Galactocentric distances are compared with the predicted values for the 
rotation curve, and the differences are attributed to the different ages of 
the evolved stars. Adopting the relation between the ages and the velocity 
dispersions determined by the Geneva-Copenhagen survey, the age distribution 
can be derived. Second, the U, V, W, velocity components of the stars are 
determined, and the corresponding age-velocity dispersion relations are 
used to infer the age distribution. These methods have been applied to two 
samples of PN in the Galaxy. The results are similar for both samples, and show 
that the age distribution of the PN central stars concentrates in ages lower 
than 5 Gyr, peaking at about 1 to 3 Gyr.}

\resumen{ }

\addkeyword{ISM: planetary nebulae: general}
\addkeyword{Stars: AGB and Post-AGB}
\addkeyword{Stars: general}
\addkeyword{Stars: fundamental parameters}

\begin{document}
\maketitle

\section{Introduction}
\label{section1}

Planetary nebulae (PN) are evolved objects ejected by stars with main sequence masses in the 
range of 0.8 and 8 $M_\odot$, so that the expected ages of their central stars are of the order 
or greater than about 1 Gyr. However, the relatively large mass bracket of their progenitor 
stars implies that an age distribution is to be expected, which has some consequences in the 
interpretation of the PN data in the Galaxy and other stellar systems. The determination of 
ages of the central stars is a difficult problem, and most usual methods have large 
uncertainties when applied to intermediate and old age objects. We have recently developed 
three different methods to estimate the age distribution of the CSPN (Maciel, Costa \& Idiart 
\citeyear{mci2010}, see also Maciel et al. \citeyear{mcu2003}, \citeyear{mlc2005}, 
\citeyear{mlc2006}), and applied these methods to a sample of PN in the  disk of the Galaxy, most 
of which are located in the solar neighbourhood, within 3 kpc from the Sun. These methods 
include the determination of the age distribution of CSPN using  (i) an age-metallicity  
relation that also depends on the Galactocentric distance, (ii) an  age-metallicity relation 
obtained for the disk,  and  (iii)  the central star masses obtained from the observed 
nitrogen abundances. We concluded that most CSPN in our sample have ages under 6 Gyr, 
and that the age distribution is peaked around 2-4 Gyr. The average uncertainties were 
estimated as 1-2 Gyr, and the results were compared with the expected distribution based 
both on the observed mass distribution of white dwarfs and on the age distribution derived 
from available masses of CSPN. 

In the present work we develop two additional and more accurate methods to estimate the age 
distribution of the CSPN based on their kinematical properties, namely:  (i) A method 
based on the expected rotation velocities of the nebulae at their Galactocentric distances,
which are then compared with the predicted values for a given rotation curve; the 
differences are attributed to the different ages of the evolved stars; (ii) A method 
based on the derived U, V, W, velocity components of the stars and their corresponding
dispersions. In both cases, the age-velocity dispersion relations from the Geneva-Copenhagen 
survey  are used to infer the age distribution. These methods are applied to two PN samples, 
(i) the previous sample of disk PN used by Maciel, Costa \& Idiart (\citeyear{mci2010}),
for which a detailed data set is available, and (ii) a sample containing all PN for which 
accurate radial velocities are known. The methods are developed in Section~2, and the samples 
used are described in Section~3. The main results and discussion are given in Section~4.

\section{Determination of the age distribution of CSPN}
\label{section2}

\subsection{Method 1: The PN rotation velocity}

As objects of intermediate age, PN the disk of the Galaxy describe a rotation curve similar 
to the one defined by younger objects, such as HII regions, although with a higher dispersion, 
as discussed in detail by Maciel and Lago (\citeyear{ml2005}). Therefore, the discrepancies 
between the rotation velocities inferred from the PN  radial velocities and distances and the 
expected velocities from the known rotation curve may be at least partially ascribed to their 
evolved status. In other words, a given nebulae located at a distance $d$, with galactic 
coordinates $\ell$ and $b$ and observed heliocentric radial velocity $V_r(hel)$ can be associated 
with a rotation velocity $\theta(R)$, after obtaining its  Galactocentric distance $R$ and its  
radial velocity relative to the Local Standard of Rest (LSR), $V_r(LSR)$. Assuming circular 
orbits, the rotation velocity $\theta(R)$  at the Galactocentric distance $R$ can be written as

   \begin{equation}
     \theta(R) = {R \over R_o} \ \biggl[{V_r(LSR) \over \sin \ell \, \cos b} + \theta_0\biggr]
   \end{equation}

\noindent
where $R_0$ and $\theta_0$ are the Galactocentric distance and rotation velocity at the solar
position (see for example Maciel \& Lago \citeyear{ml2005}, Maciel \& Dutra \citeyear{md1992}).
On the other hand, the expected rotation velocity at the given Galactocentric distance, 
$\theta_c(R)$, can be obtained from an adopted rotation curve. The difference $\Delta \theta = 
\vert{\theta(R) - \theta_c(R)}\vert$  can then be considered as proportional to the age difference 
between the PN and the objects defining the rotation curve. We have adopted the radial velocities 
from the catalogue by Durand et al. (\citeyear{durand}), and two distance scales, those by Maciel 
(\citeyear{m1984}) and Stanghellini et al. (\citeyear{ssv2008}). The first one was based on a 
relationship between the ionized mass and the radius of the nebulae, while the second is an update 
of the distance scale by Cahn et al. (\citeyear{cks1992}), using a modified Shklovksy method 
following Daub (\citeyear{daub}). Since the distances of planetary nebulae in the Galaxy may contain 
large individual uncertainties, the use of two different scales which are considered as 
\lq\lq short\rq\rq\ (Maciel \citeyear{m1984}) and \lq\lq long\rq\rq (Stanghellini et al. 
\citeyear{ssv2008}) warrants that these uncertainties will not affect the derived age distributions. 
We have adopted $R_0 = 8.0\,$kpc for the distance of the Sun to the centre and 
$\theta_0 = 220\,$km/s for the rotation velocity at $R_0$. Slightly different values can be found 
in the literature (see for example Perryman, \citeyear{perryman}, and Reid, \citeyear{reid}), 
but the values above are frequently adopted, so that a comparison with other works is made easier. 
For the \lq\lq theoretical\rq\rq\ rotation curve we have also adopted two possibilities, namely, 
the PN derived curve by Maciel \& Lago (\citeyear{ml2005}), and the HII region derived curve by 
Clemens (\citeyear{clemens}). In the first case, the rotation velocity can be written as

   \begin{equation}
     \theta_c(R) = a_0 + a_1 \, R + a_2 \, R^2  \ .
   \end{equation}

\noindent
where the constants are $a_0 = 269.2549$, $a_1 = -14.7321$, and $a_2 = 0.7847$,  the Galactocentric 
distance $R$ is in kpc and $\theta_c(R)$ in km/s.  For the CO/HII region based Clemens (\citeyear{clemens})
curve, we have made an adjustment for $R_0 = 8.0\,$kpc and $\theta_0(R) = 220\,$km/s, in which 
case we have 

   \begin{equation}
     \theta_c(R) = \sum a_i\, R^i  \ .
   \end{equation}

\noindent
where the constants are given in Table 1, with the same units as in Eq.(2).

\begin{table*}
\small
\caption[]{Coefficients of the polynomial given by Eq. (3).}
\label{table1}
\begin{flushleft}
\begin{tabular}{cccccc}
\noalign{\smallskip}
\hline\noalign{\smallskip}
$R$ (kpc) & $0-0.765$ & $0.765-2.9$ & $2.9-3.825$ & $3.825-13$ & $> 13$ \\
\noalign{\smallskip}
\hline\noalign{\smallskip}
$a_0$ & 0.0         & 325.0912     & 329.8        & $-$2346.0      & 230.6 \\
$a_1$ & 3069.81     & $-$248.1467  & $-$250.1     & 2507.60391     &  $-$  \\
$a_2$ &  $-$15809.8 & 231.87099    & 231.87099    & $-$1024.068760 &  $-$  \\
$a_3$ &  43980.1    & $-$110.73531 & $-$110.73531 & 224.562732     &  $-$  \\
$a_4$ & $-$68287.3  & 25.073006    & 25.073006    & $-$28.4080026  &  $-$  \\
$a_5$ & 54904.0     & $-$2.110625  & $-$2.110625  & 2.0697271      &  $-$  \\
$a_6$ & $-$17731.0  &  $-$         &   $-$        & $-$0.080508084 &  $-$  \\
$a_7$ &  $-$        &  $-$         &   $-$        & 0.00129348     &  $-$  \\
\noalign{\smallskip}
\hline
\end{tabular}
\end{flushleft}
\end{table*}

The recent Geneva-Copenhagen Survey of the Solar Neighbourhood (cf. Nordstr\"om et al. 
\citeyear{nordstrom}, Holmberg et al. \citeyear{holmberg2007}, \citeyear{holmberg2009}) 
has considerably improved the relations involving the ages, kinematics, and chemical
composition of a large sample containing about 14000 F and G nearby stars. Using basically  
the original {\it Hipparcos} parallaxes, uvby$\beta$\ photometry and the Padova stellar
evolution models, several basic relations were investigated. In particular,  high 
correlations have been obtained between the velocity dispersions $\sigma_U$, $\sigma_V$, 
$\sigma_W$, and $\sigma_T$  and the age of the star, which clearly show a smooth
increase of the velocity dispersions in the U, V, W components and total velocity $T$
with time. From the calibration by Holmberg et al. (\citeyear{holmberg2009}) these 
correlations can be approximately written as

   \begin{equation}
     \log \sigma = a \, \log t + b \ ,
   \end{equation}

\noindent
where the age $t$ is in Gyr and the constants $a$, $b$ are given in Table~2. This approximation
is valid in the age interval $0 < t {\rm (Gyr)} < 14$ with an estimated average age uncertainty 
of about 25\%. Method~1 consists of assuming that the discrepancy in the rotation velocity 
$\Delta \theta$ is due to the evolved status of the CSPN, so that we should expect a correlation 
between $\Delta \theta$ and the velocity dispersion, as given by Eq.~(4). Since in this method 
we are using the rotation velocity, we have considered two possibilities, according to which the 
velocity discrepancy $\Delta \theta$ can be associated with (i) the $V$ component of the total 
velocity ($\sigma_V$), or (ii) the total velocity ($\sigma_T$). Moreover, since we are adopting 
two distance scales and two theoretical rotation curves, we have 8 different age distributions 
for Method~1, characterized by the timescales $t_1$ to $t_8$, as explained in Table~3.

\begin{table*}
\small
\caption[]{Coefficients of Eq. (4).}
\begin{changemargin}{4cm}{4cm}
\label{table2}
\begin{flushleft}
\begin{tabular}{ccc}
\noalign{\smallskip}
\hline\noalign{\smallskip}
	& $a$ & $b$ \\
\noalign{\smallskip}
\hline\noalign{\smallskip}
$U$   & 0.39 & 1.31 \\
$V$   & 0.40 & 1.10 \\
$W$   & 0.53 & 0.94 \\
Total & 0.40 & 1.40 \\
\noalign{\smallskip}
\hline
\end{tabular}
\end{flushleft}
\end{changemargin}
\end{table*}

\begin{table*}
\small
\caption[]{Parameters for Method 1.}
\begin{changemargin}{2cm}{2cm}
\label{table3}
\begin{flushleft}
\begin{tabular}{cccc}
\noalign{\smallskip}
\hline\noalign{\smallskip}
Distance & Rotation Curve & Dispersion & Age \\
\noalign{\smallskip}
\hline\noalign{\smallskip}
Maciel       & PN      & $\sigma_V$ & $t_1$ \\
Maciel       & PN      & $\sigma_T$ & $t_2$ \\
Maciel       & Clemens & $\sigma_V$ & $t_3$ \\
Maciel       & Clemens & $\sigma_T$ & $t_4$ \\
Stanghellini & PN      & $\sigma_V$ & $t_5$ \\
Stanghellini & PN      & $\sigma_T$ & $t_6$ \\
Stanghellini & Clemens & $\sigma_V$ & $t_7$ \\
Stanghellini & Clemens & $\sigma_T$ & $t_8$ \\
\noalign{\smallskip}
\hline
\end{tabular}
\end{flushleft}
\end{changemargin}
\end{table*}

\subsection{Method 2: The U, V, W velocity components}

Method~2 is also a kinematic method, and in principle more accurate than Method~1, 
as discussed in more detail in Section~4. 
From the PN radial velocities and distances, we have estimated their proper motions both in 
right ascension and declination, $\mu_\alpha$ and $\mu_\delta$. We have assumed that, in average, 
the tangential velocities are similar to the radial velocities, namely $V_t \simeq V_r$. In 
view of the large distances of the nebulae, this hypothesis in practice does not introduce any 
major uncertainties in the results. Considering further the equatorial coordinates 
($\alpha,\delta$) of the PN, we have used the equations by Boesgaard and Tripicco 
(\citeyear{boesgaard}) to derive the $U$, $V$, $W$ velocity components of the nebulae, as well 
as the total velocity $T$ and the velocity dispersions $\sigma_U$, $\sigma_V$, $\sigma_W$, and 
$\sigma_T$. According to these equations we derive the following parameters: $C = f(d)$,
$X = f(C, \mu_\alpha, \mu_\delta, \alpha, \delta, V_r)$,
$Y = f(C, \mu_\alpha, \mu_\delta, \alpha, \delta, V_r)$, and
$Z = f(C, \mu_\delta, \delta, V_r)$, from which the velocities can be written as
$U = f(X, Y, Z)$, $V = f(X, Y, Z)$, $W = f(X, Y, Z)$, and $T = f(X, Y, Z)$,
so that the dispersions are given by

   \begin{equation}
     \sigma_i = \sqrt{(V_i-\bar V_i)^2}
   \end{equation}

\noindent
where $V_i$ stands for the velocities $U, V, W, T$. Then, we have again used the detailed 
correlations between the velocity dispersions and the ages as given by the Geneva-Copenhagen 
survey (Holmberg et al. \citeyear{holmberg2009}), adopting the same coefficients given in 
Table~2. We have used the same distance scales (Maciel \citeyear{m1984} and Stanghellini et 
al. \citeyear{ssv2008}), so that we have again 8 different age distributions, corresponding 
to the timescales $t_9$ to $t_{16}$, as described in Table~4.

\begin{table*}
\small
\caption[]{Parameters for Method 2.}
\begin{changemargin}{3.5cm}{3.5cm}
\label{table4}
\begin{flushleft}
\begin{tabular}{ccc}
\noalign{\smallskip}
\hline\noalign{\smallskip}
Distance & Dispersion & Age \\
\noalign{\smallskip}
\hline\noalign{\smallskip}
Maciel & $\sigma_U$ & $t_9$ \\
Maciel & $\sigma_V$ & $t_{10}$ \\
Maciel & $\sigma_W$ & $t_{11}$ \\
Maciel & $\sigma_T$ & $t_{12}$ \\
Stanghellini & $\sigma_U$ & $t_{13}$ \\
Stanghellini & $\sigma_V$ & $t_{14}$ \\
Stanghellini & $\sigma_W$ & $t_{15}$ \\
Stanghellini & $\sigma_T$ & $t_{16}$ \\
\noalign{\smallskip}
\hline
\end{tabular}
\end{flushleft}
\end{changemargin}
\end{table*}

In practice, we have considered several additional cases, in order to better investigate
the hypothesis of $V_t \simeq V_r$. Assuming that these velocities are of the same magnitude, 
but allowing for the possibility of different signs, we have as a result several possibilities 
for the proper motions $\mu_\alpha$ and $\mu_\delta$, all of which are consistent with either 
$V_t \simeq V_r$ or $\vert V_t\vert \simeq \vert V_r\vert$. It turns out that these 
possibilities produce very similar age distributions, which will be discussed in 
Section~4. Therefore, we will present only the distributions of the ages $t_9$ to $t_{16}$, 
as defined in Table~4, for the cases where $\mu_\alpha \simeq \mu_\delta \simeq 0$.

An interesting alternative to overcome the lack of proper motion and 
tangential velocity measurements would be to apply the singular value decomposition
(SVD) technique, as used by Branham (\citeyear{branham}) to solve the inverse
problem, that is, obtaining the space velocities from available proper motions. However,
in view of the similarity of the results for different assumptions regarding
the tangential velocities, it is unlikely that this technique would produce
very different results than presented here.

\section{The  samples}
\label{section3}

As mentioned in the Introduction, we have considered two samples of Milky Way PN. In order to make 
comparisons with our previous work, we have first considered the same sample used by Maciel et al. 
(\citeyear{mcu2003}, \citeyear{mlc2005}, \citeyear{mlc2006}), which we will call Sample~1. This 
sample contains 234 well-observed nebulae located in the solar neighbourhood and in the  
disk, for which all data were obtained with the highest accuracy. Their Galactocentric distances 
are in the range $4 <  R {\rm (kpc)} < 14$, and most (69\%) are located in the solar neighbourhood, 
with distances $d < 3\,$kpc.

The second sample considered in this work, called Sample~2, includes all the nebulae for which 
accurate radial velocities are available in the catalogue by Durand et al. (\citeyear{durand}), 
comprising 867 objects. This is a more complete sample, so that it is expected that the derived 
results can be extended to the observed population of PN in the Galaxy. In both samples, the number of 
nebulae used depends on the availability of the statistical distances. The actual number of objects 
from the  Maciel (\citeyear{m1984}) and Stanghellini et al. (\citeyear{ssv2008}) distance 
scales are 195 and 170 for Sample 1 and 493 and 403 for Sample~2, respectively. We have then applied 
the approximation given by Eq. (4) for both samples, with the coefficients shown in Table~2,
considering only the objects for which ages in the interval $0 < t ({\rm Gyr})  < 14$ could be 
obtained.

\section{Results and discussion}
\label{section4}

The main results for the age distribution of the CSPN are shown in Figures 1-4, where we have
used the age parameter definitions given in Tables~3 and 4 for Methods 1 and 2, respectively. 
Figures~1 and 2 refer to Sample~1, while figures~3 and 4 refer to Sample~2. It can be seen
that the age distributions obtained by both methods are similar, in the sense that most
objects have ages under 5 Gyr, with a strong peak at ages typically between 1 and 3 Gyr.
The histograms of Figures 3-4 are summarized in Table~5, where the fraction of stars
obtained by Method~1 (ages $t_1$ to $t_8$) and Method~2 (ages $t_9$ to $t_{16}$) are shown
for three age bins, namely $0 - 3\,$Gyr, $3 - 6\,$Gyr, and $t > 6\,$Gyr.

\begin{table*}
\small
\caption[]{Fraction of stars at three age intervals.}
\begin{changemargin}{2cm}{2cm}
\label{table5}
\begin{flushleft}
\begin{tabular}{ccccc}
\noalign{\smallskip}
\hline\noalign{\smallskip}
 & $\Delta t$ (Gyr) & $0 - 3$ & $3 - 6$  & $> 6$ \\
\noalign{\smallskip}
\hline\noalign{\smallskip}
Method 1 & $t_1$    & $0.57$ & $0.13$ & $0.30$  \\
         & $t_2$    & $0.62$ & $0.18$ & $0.20$  \\
         & $t_3$    & $0.57$ & $0.19$ & $0.24$  \\
         & $t_4$    & $0.67$ & $0.18$ & $0.16$  \\
         & $t_5$    & $0.51$ & $0.13$ & $0.36$  \\
         & $t_6$    & $0.71$ & $0.17$ & $0.12$  \\
         & $t_7$    & $0.61$ & $0.15$ & $0.24$  \\
         & $t_8$    & $0.71$ & $0.11$ & $0.18$  \\
         &          &        &        &         \\
Method 2 & $t_9$    & $0.76$ & $0.12$ & $0.12$  \\
         & $t_{10}$ & $0.79$ & $0.10$ & $0.11$  \\
         & $t_{11}$ & $0.92$ & $0.04$ & $0.04$  \\
         & $t_{12}$ & $0.77$ & $0.18$ & $0.05$  \\
         & $t_{13}$ & $0.78$ & $0.10$ & $0.12$  \\
         & $t_{14}$ & $0.78$ & $0.11$ & $0.11$  \\
         & $t_{15}$ & $0.93$ & $0.03$ & $0.04$  \\
         & $t_{16}$ & $0.76$ & $0.18$ & $0.06$  \\
\noalign{\smallskip}
\hline
\end{tabular}
\end{flushleft}
\end{changemargin}
\end{table*}

The similarity
of the results of both methods is remarkable, especially considering that Method~2
is probably more accurate than Method~1. Method~2 consists of straightforward calculations
of the velocities and velocity dispersions followed by an application of relatively
accurate correlations involving the kinematics and ages of the objects considered. On
the other hand, Method~1 is based on the assumption that the differences between the observed
and predicted rotation velocities are essentially due to age effects. However, other processes
may be important, such as deviations from the circular rotation, which is particularly important
for nearby objects.  According to Table~5, in all cases the vast majority of CSPN have ages 
under 3 Gyr. For Method~1 the total fraction of objects with $t \leq 3\,$ Gyr is $50 - 70\%$, 
while for Method~2 this fraction is somewhat higher, $70 - 90\%$. It is unlikely that this is a 
result from biased samples, as the results for the larger Sample~2 are essentially the same as 
in the smaller Sample~1. It should be pointed out that the latter, albeit smaller, includes 
only well studied nebulae, for which all individual parameters (distances, velocities, abundances) 
are better determined. 

Also, there are no significant differences in the results using the different 
velocity components $U$, $V$, $W$, and $T$. For Method~1, the distributions 
using the $V$ velocity component are essentially the same as those using the total 
velocity, for both distance scales and samples. For Method~2, the distributions
are slightly more concentrated in the first few age bins for the $W$ component, compared
with the distributions for the $U$ and $V$ components and the  total velocity, again
for both distance scales and samples. Since the $W$ component is more clearly associated
with the disk heating, essentially caused by age effects, the corresponding distributions
are probably more accurate. 

Similar remarks can be made regarding the adopted values for the proper motions. As 
mentioned at the end of Section~2, the results shown here assume that 
$\mu_\alpha \simeq \mu_\delta \simeq 0$. Adopting nonzero values for these
quantities ($\mu_\alpha \simeq \mu_\delta \neq 0$), either the $V$ or $W$ component 
distributions become slightly less concentrated at the first few age bins, but most 
objects still have ages under about 4~Gyr. Again, the application of
the SVD technique could be useful to confirm these results.

The uncertainties in the distances of the Milky Way PN are difficult to estimate, but
the procedure adopted here ensures that the obtained age distributions are not 
particularly affected by the individual distances of the objects in the samples.
As mentioned in Section~2, we have adopted two very different statistical scales,
and the derived age distrbutions are essentially the same in both cases. The
individual distances may depend on the particular scale, but the results shown
in Figures~1--4 and in Table~5 do not depend on the choice of the distance scale.
This can be seen by comparing the results for the timescales $t_1 - t_4$ with those
for $t_5 - t_8$, or the results for $t_9 - t_{12}$ with those for $t_{13} - t_{16}$.

The uncertainties in the radial velocities also do not seem to have an important
effect on the age distributions. In the catalogue by Durand et al. (\citeyear{durand}),
most objects ($\sim 90\%$) have uncertainties smaller than 20 km/s, and many
objects have much lower uncertainties. Concerning Method~1, from Maciel \& Lago
(\citeyear{ml2005}), the average rms deviation in the rotation velocity is 
about 50 km/s for PN, which can be compared with the values of about
20 km/s for HII regions (see also Clemens \citeyear{clemens} and Maciel \& Dutra
\citeyear{md1992}). 

Probably the main uncertainty of the age distributions is due to the calibration 
between the stellar ages and the velocity dispersions, given by Eq. (4), which affects
both Method~1 and 2. From the Geneva-Copenhagen Survey, this relation has a 
dispersion of about 20 km/s in average, which corresponds roughly to an age
uncertainty of about 25\%, amounting to less than 1.2 Gyr for the objects of
Figures~1--4. Therefore, the uncertainties of the present method are comparable
and probably smaller than in the case of the methods based on age-metallicity
relations considered by Maciel et al. (\citeyear{mci2010}).

The results for Sample~2 are not essentially different from those of Sample~1, so that
a direct comparison can be made with the results by Maciel et al. (\citeyear{mci2010}).
The results of both investigations are similar, even though the present methods are 
completely independent of the metallicity-based methods used by Maciel et al. (\citeyear{mci2010}).
The main difference is that the kinematic methods used in the present investigation suggest
somewhat lower ages for the CSPN in our samples. In this respect, these results fit
nicely to the probability distribution for the progenitors of the CSPN according to
Maciel et al. (\citeyear{mci2010}, cf. figure~7, dashed line). In this case the well
known relation between the main sequence mass and the stellar ages  by Bahcall \& Piran 
(\citeyear{bahcall}) was adopted, taking $t = 10\,$Gyr for $1 M_\odot$ stars on the 
main sequence. Taking into account the uncertainties of the methods, which are 
typically in the range $1-2\,$ Gyr, this case was considered as the most realistic,
so that it is reassuring that the kinematic methods produce similar results.

   \begin{figure}
   \centering
   \includegraphics[angle=-90, width=13.5cm]{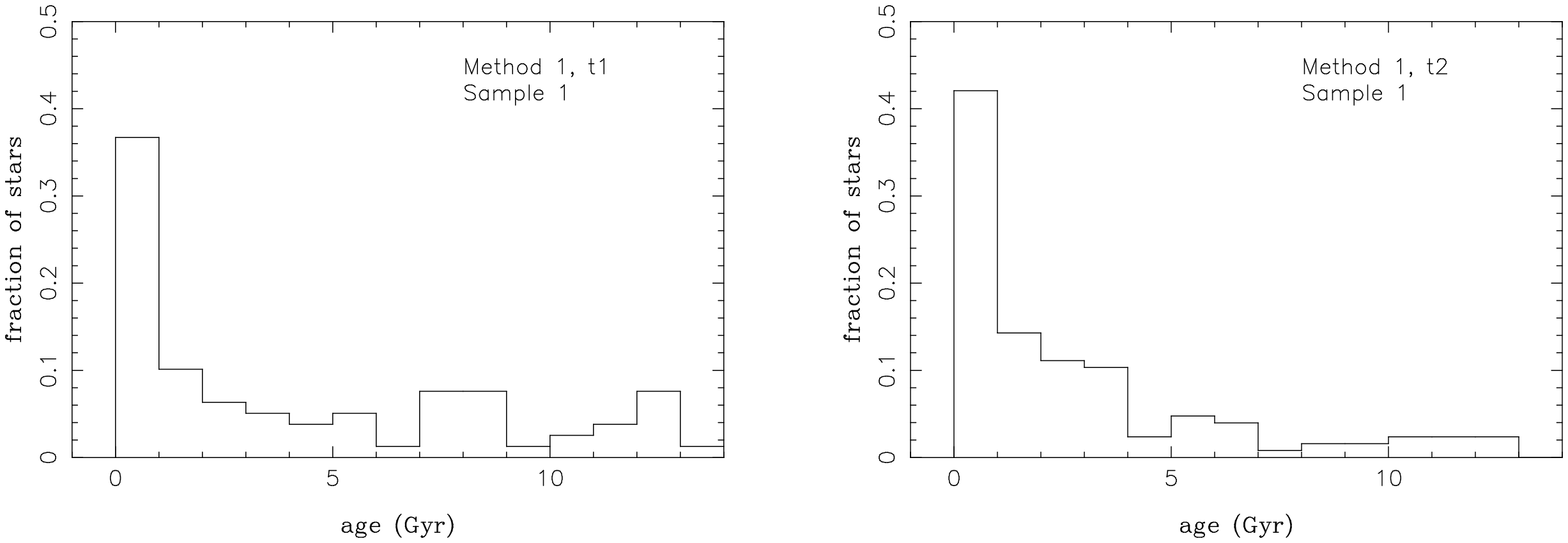}
   \vskip 0.6 true cm
   \includegraphics[angle=-90, width=13.5cm]{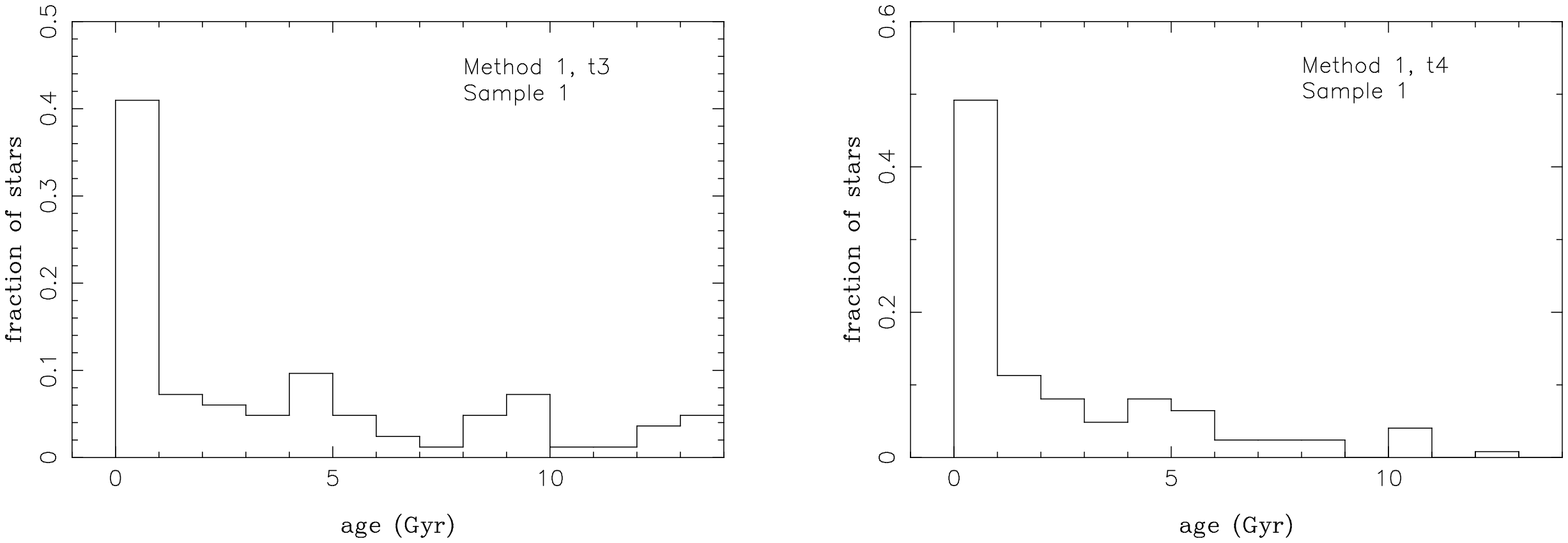}
   \vskip 0.6 true cm
   \includegraphics[angle=-90, width=13.5cm]{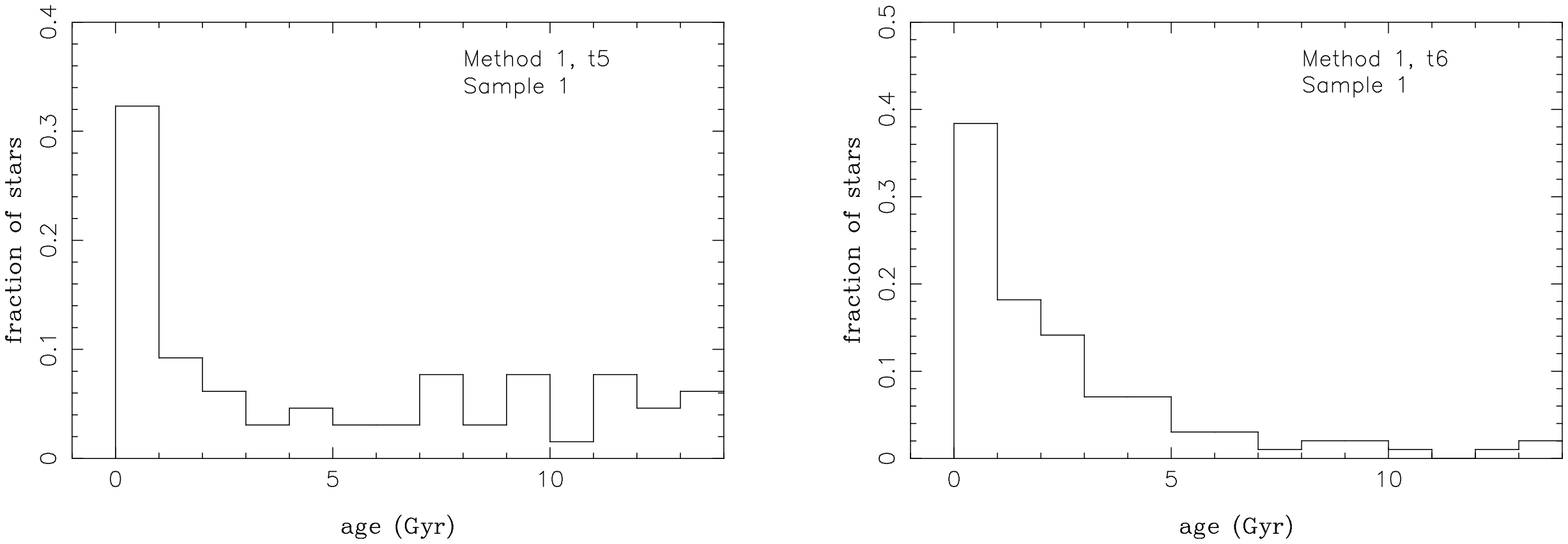}
   \vskip 0.6 true cm
   \includegraphics[angle=-90, width=13.5cm]{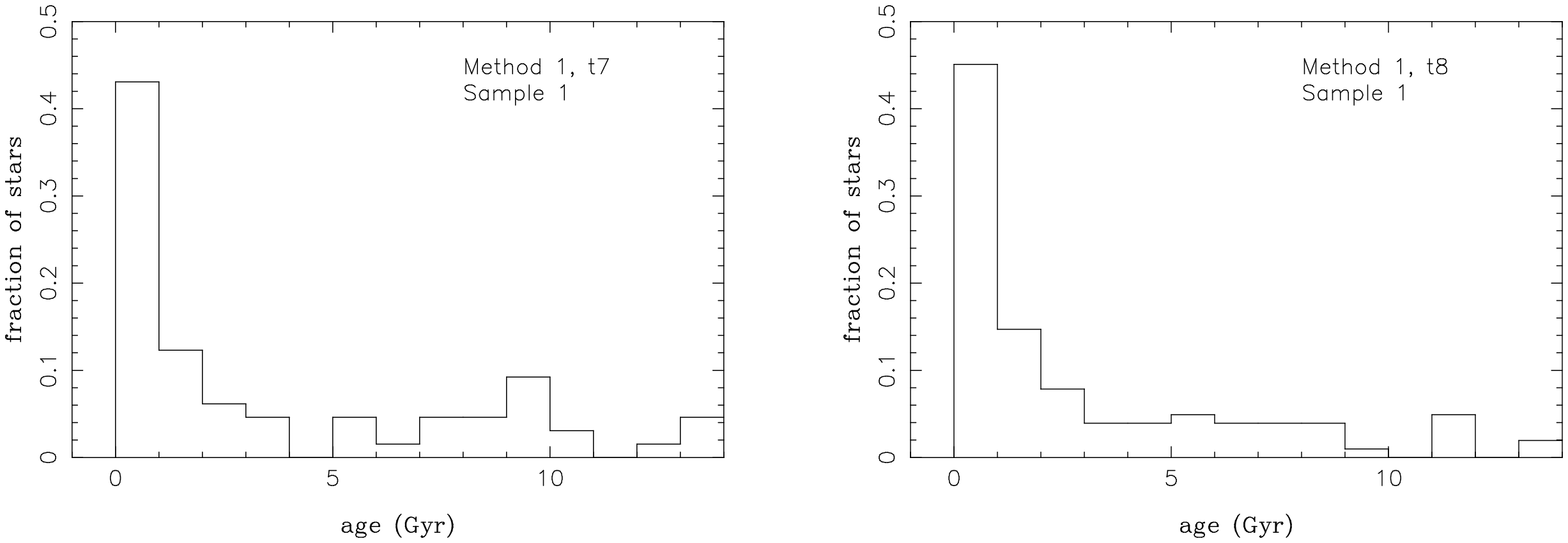}
      \caption{Age distribution of CSPN, Method 1, Sample 1.}
   \label{fig1}
   \end{figure}


   \begin{figure}
   \centering
   \includegraphics[angle=-90, width=13.5cm]{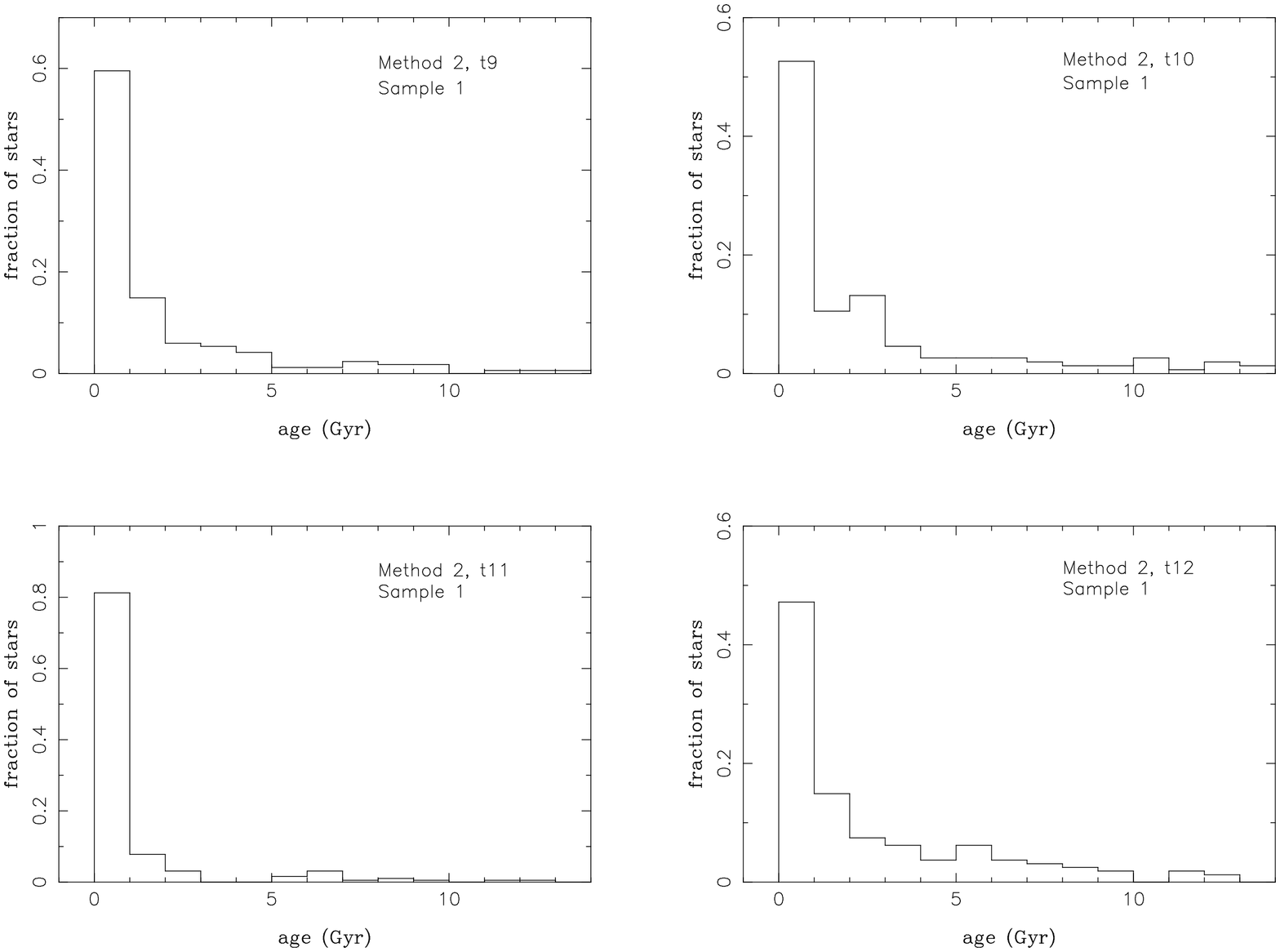}
   \includegraphics[angle=-90, width=13.5cm]{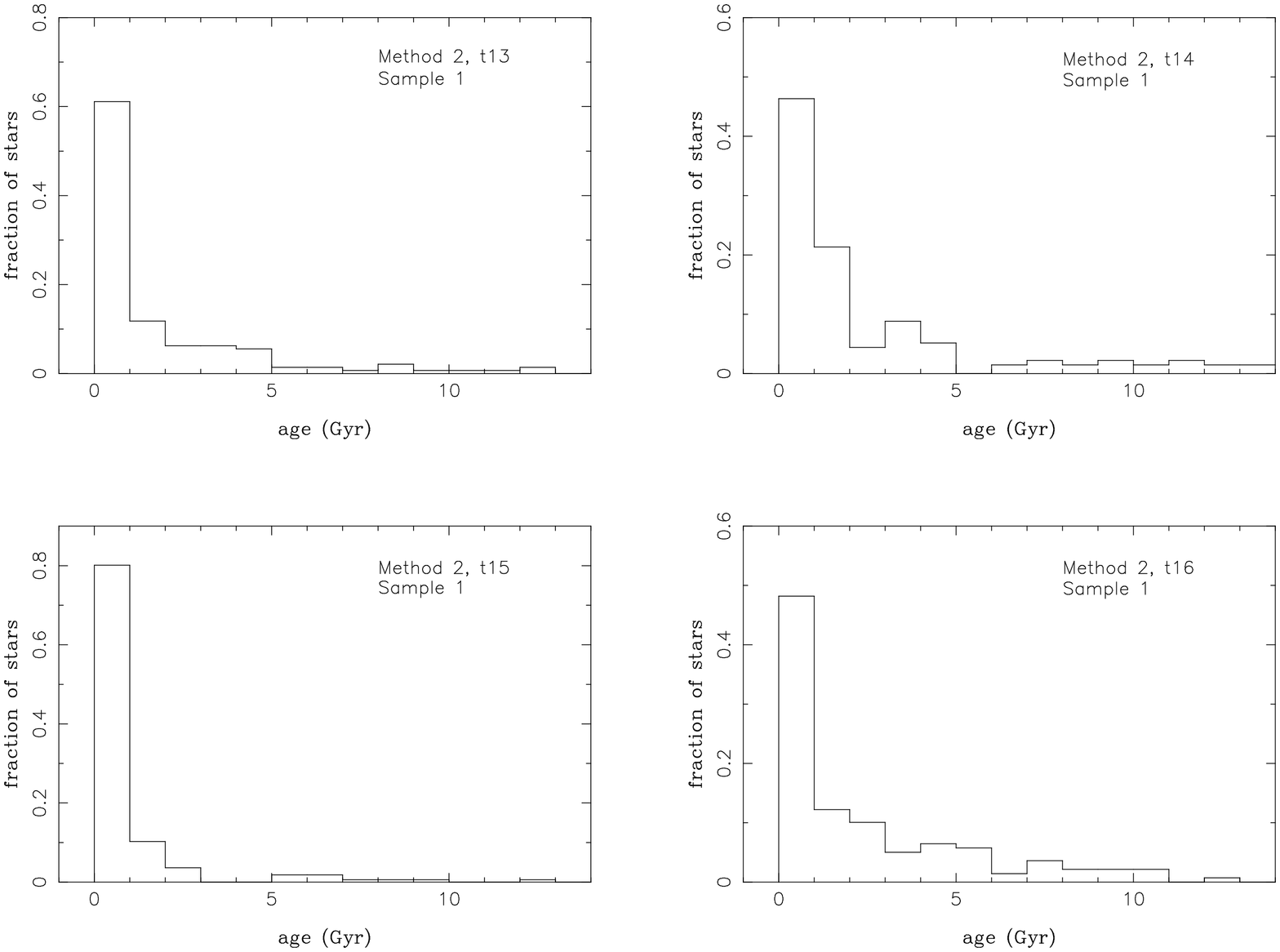}
      \caption{Age distribution of CSPN, Method 2, Sample 1.}
   \label{fig2}
   \end{figure}


   \begin{figure}
   \centering
   \includegraphics[angle=-90, width=13.5cm]{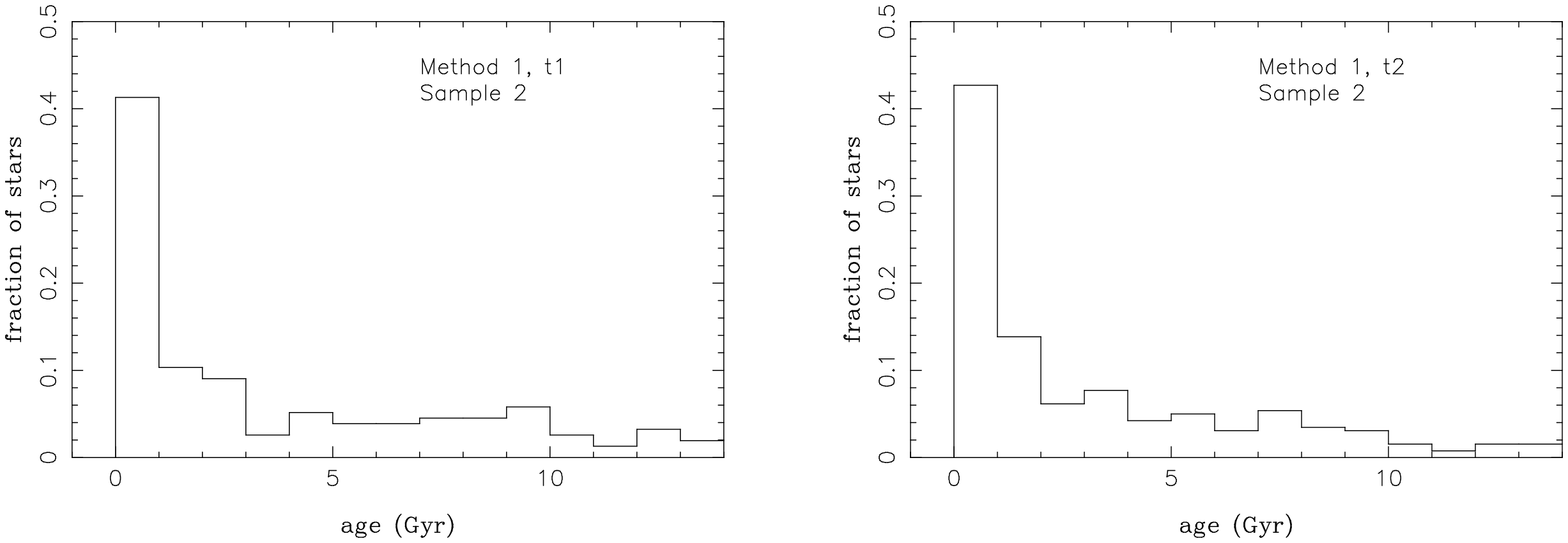}
   \vskip 0.6 true cm
   \includegraphics[angle=-90, width=13.5cm]{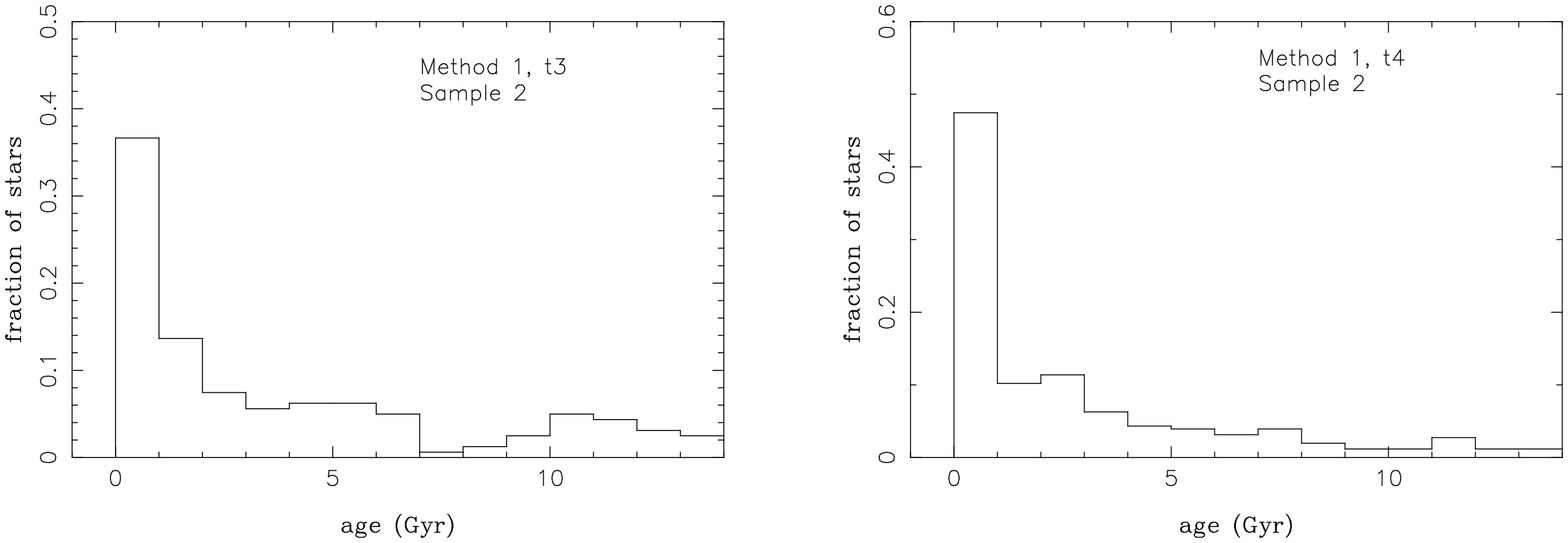}
   \vskip 0.6 true cm
   \includegraphics[angle=-90, width=13.5cm]{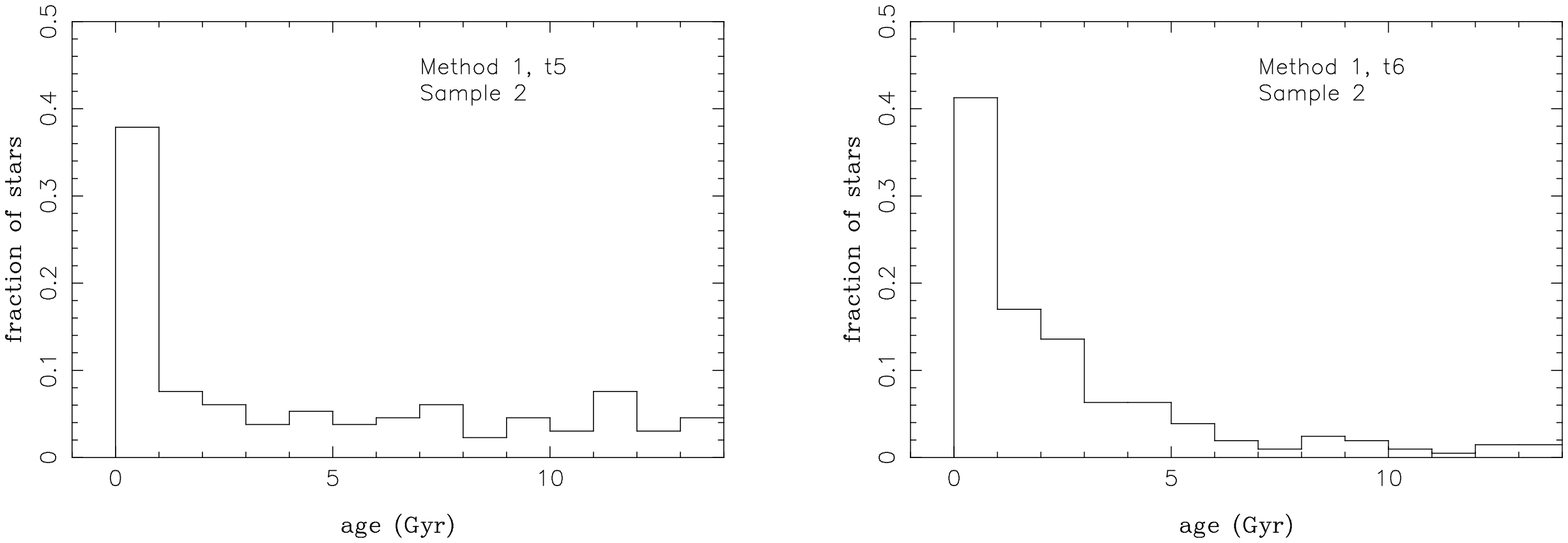}
   \vskip 0.6 true cm
   \includegraphics[angle=-90, width=13.5cm]{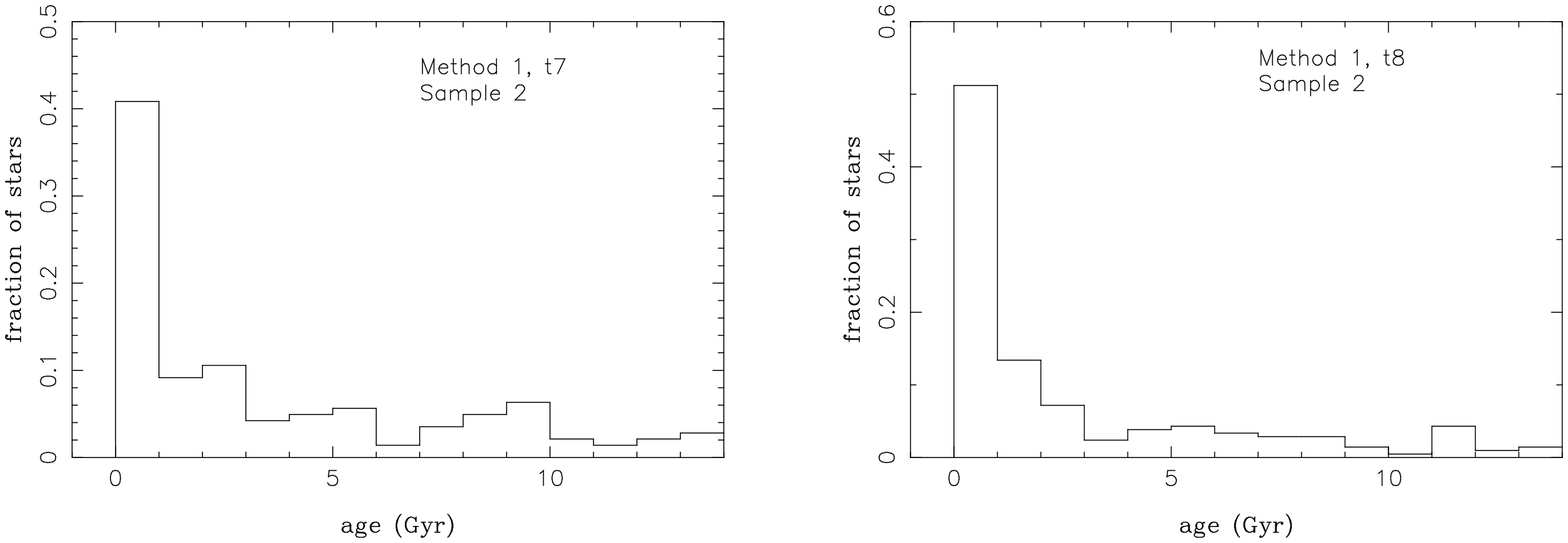}
      \caption{Age distribution of CSPN, Method 1, Sample 2.}
   \label{fig3}
   \end{figure}


   \begin{figure}
   \centering
   \includegraphics[angle=-90, width=13.5cm]{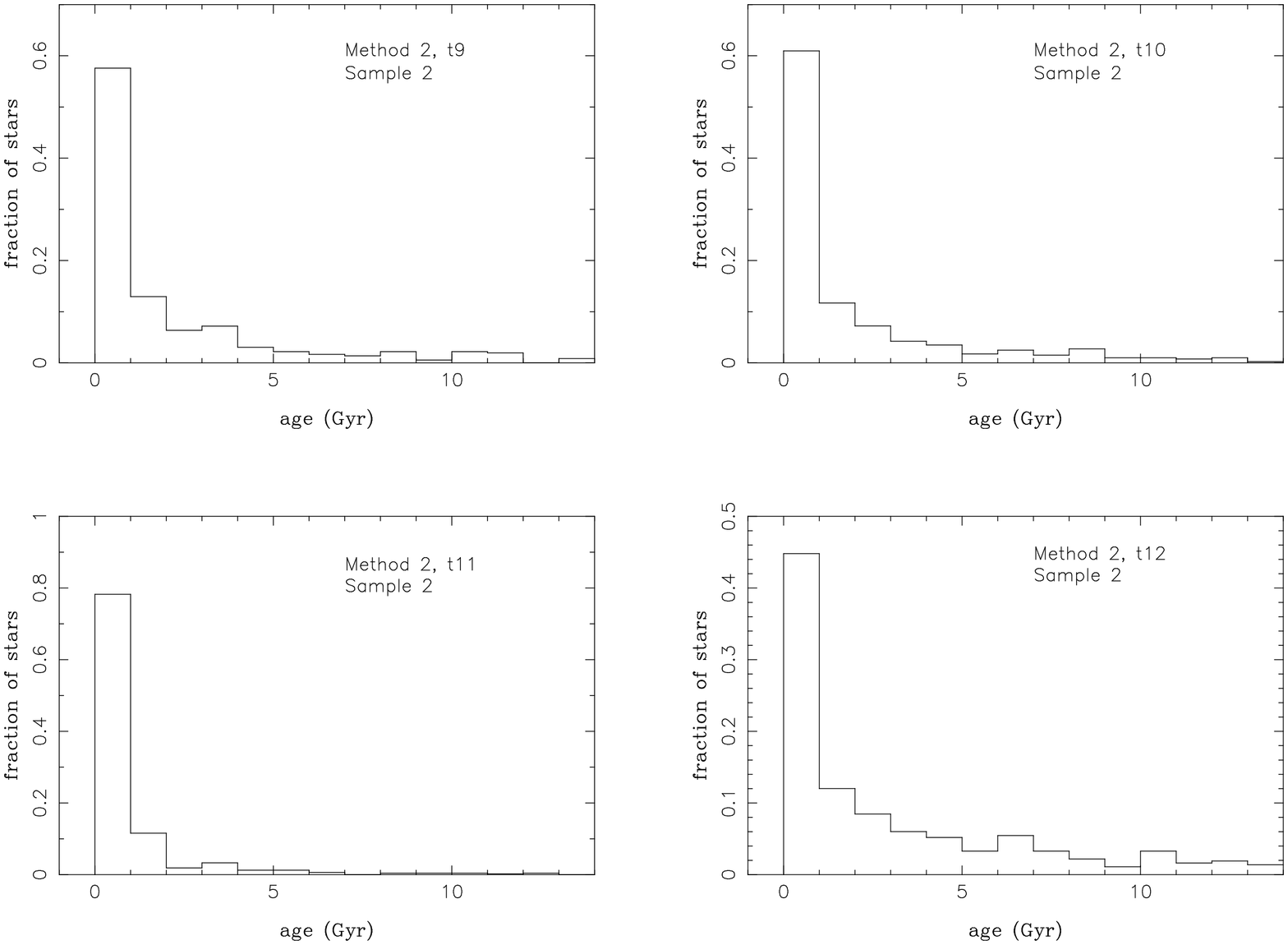}
   \includegraphics[angle=-90, width=13.5cm]{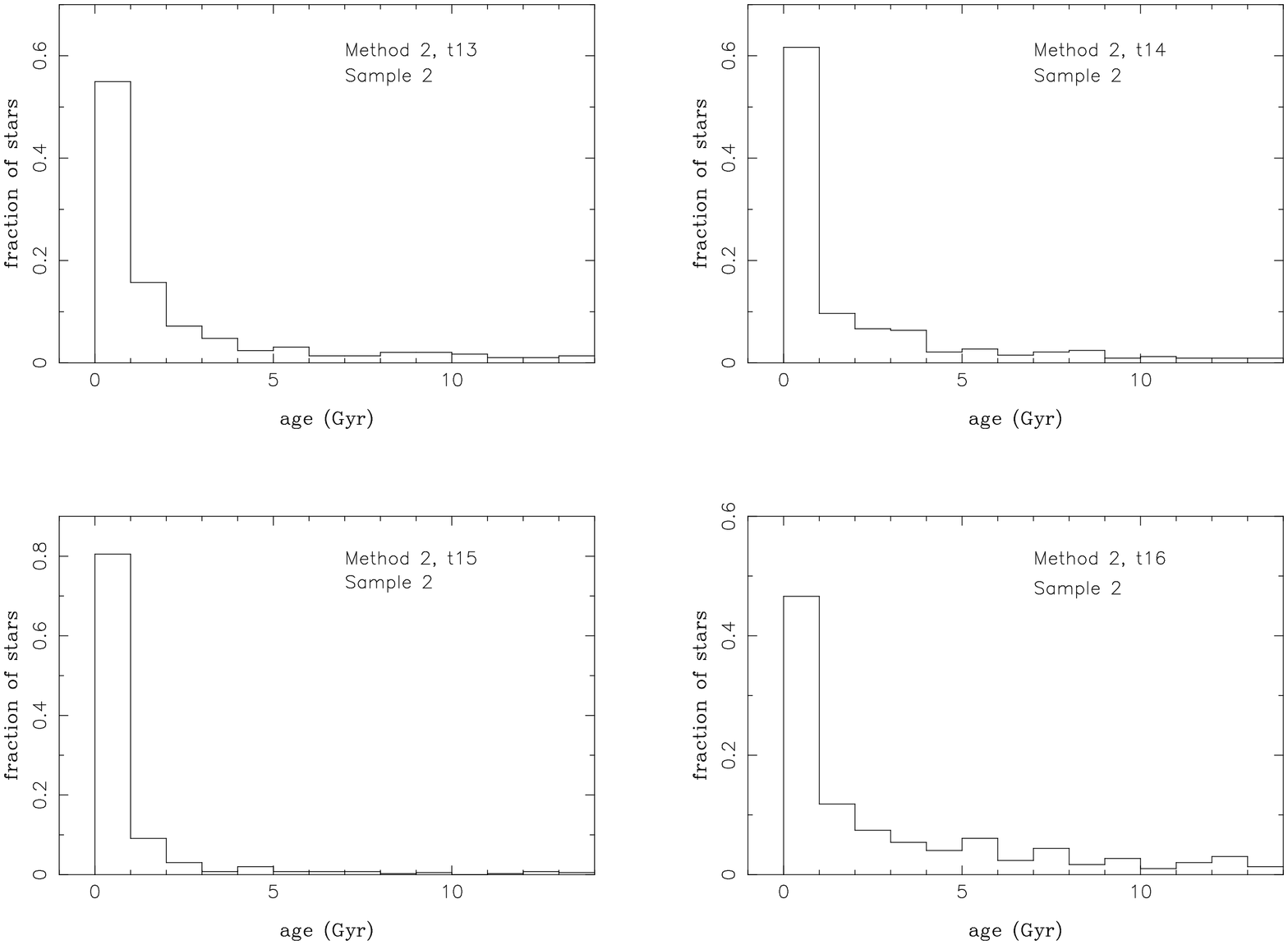}
      \caption{Age distribution of CSPN, Method 2, Sample 2.}
   \label{fig4}
   \end{figure}


\bigskip
{\it Acknowledgements. We thank Dr. R. Branham, Jr., for some interesting comments on
an earlier version of this paper. This work was partly supported by FAPESP and CNPq.}

\end{document}